\documentclass[12pt,preprint]{aastex}

\shorttitle{Acceleration of solar-type winds}
\shortauthors{Zouganelis et al.}

\begin{document}

\title{Acceleration of weakly collisional solar-type winds}

\author{I. Zouganelis\altaffilmark{1}, N. Meyer-Vernet\altaffilmark{1}, S. Landi\altaffilmark{2}, M. Maksimovic\altaffilmark{1} and F. Pantellini\altaffilmark{1}}

\altaffiltext{1}{LESIA, Observatoire de Paris, 5 place Jules
Janssen, 92195 Meudon, France; ioannis.zouganelis@obspm.fr}
\altaffiltext{2}{Dipartimento di Astronomia e Scienza dello
Spazio, Largo Enrico Fermi 2, 50125 Firenze, Italy}

\begin{abstract}
One of the basic properties of the solar wind, that is the high
speed of the fast wind, is still not satisfactorily explained.
This is mainly due to the theoretical difficulty of treating
weakly collisional plasmas. The fluid approach implies that the
medium is collision dominated and that the particle velocity
distributions are close to Maxwellians. However the electron
velocity distributions observed in the solar wind depart
significantly from Maxwellians. Recent kinetic collisionless
models (called exospheric) using velocity distributions with a
suprathermal tail have been able to reproduce the high speeds of
the fast solar wind. In this letter we present new developments of
these models by generalizing them over a large range of corona
conditions. We also present new results obtained by numerical
simulations that include collisions. Both approaches calculate the
heat flux self-consistently without any assumption on the energy
transport. We show that both approaches - the exospheric and the
collisional one - yield a similar variation of the wind speed with
the basic parameters of the problem; both produce a fast wind
speed if the coronal electron distribution has a suprathermal
tail. This suggests that exospheric models contain the necessary
ingredients for the powering of a transonic stellar wind,
including the fast solar one.
\end{abstract}

\keywords{methods: numerical --- stars: winds, outflows --- solar
wind --- Sun: corona}

\section{Introduction} \label{introduction}

In spite of the success of the fluid models in explaining the
supersonic solar wind, it is still not known how the fast wind is
accelerated to speeds up to 800 km s$^{-1}$ and how the energy is
transported. This is mainly due to a major theoretical difficulty
of treating such weakly collisional plasmas. In fact, the Knudsen
number, which is defined as the ratio of the particle mean free
path to the density scale height, is close to unity at Earth's
orbit and larger than $10^{-3}$ in the fast wind acceleration
region. In this case, the classical heat conduction formulation
\citep{spi53} breaks down \citep{sho83}, and the low level closing
of the infinite hierarchy of MHD equations is hard to justify (see
\citet{parks04}).

Furthermore the electron velocity distribution functions (VDFs) in
the solar wind are not Maxwellians. They exhibit high energy
(nonthermal) tails that have been modeled by a halo Maxwellian
population \citep{fel75} or, more recently, by the power-law part
of a generalized Lorentzian or Kappa function \citep{mak97b}.
These tails can develop even in moderately collisional plasmas as
a result of the rapid increase of the particle free paths with
speed $(\propto v^{4})$. The existence of such electron VDFs in
the upper chromosphere has been suggested to be the reason for the
rapid rising of the temperature in the chromosphere-corona
transition region through the mechanism of gravitational velocity
filtration \citep{scu92}. Indeed, there is an increasing amount of
observational evidence showing that nonthermal VDFs may exist in
the corona and even in the high chromosphere \citep{owo99, pin99,
ess00, chi04, doy04}. Some theoretical works on the possible
generation mechanisms of such nonthermal electron distributions in
the chromosphere \citep{rob98, vin00} and the corona \citep{voc03}
have been published. Others have been trying to show that Kappa
distributions can be a natural, and quite general, state of weakly
collisional plasmas and not merely a convenient mathematical way
of describing non thermal VDFs \citep{col93, ma99, tre99, leu02,
col04}.

Investigating the effects of nonthermal VDFs in the corona
requires a kinetic approach. The simplest one is the exospheric
model that has recently provided transonic wind solutions using
non-Maxwellian VDFs for the electrons \citep{zou04}. However, the
collisionless assumption may appear as a strong intrinsic
limitation of these models. Kinetic simulations, taking into
account binary collisions between particles \citep{lan03}, suggest
that collisions might be an important ingredient for accelerating
the wind to supersonic speeds, even though this latter work does
not consider nonthermal electron VDFs. We should note that these
models do neither include the effects of plasma instabilities nor
any kind of wave-particle interaction that are sometimes invoked
to be a fundamental ingredient in the wind acceleration process
(see e.g. \citet{lie01}). Given that the real importance of these
effects is not yet established, they will be neglected in this
work.

In this letter we present new simulations based on \citet{lan03}
using Kappa distributions for the electrons and compare the
results to those of exospheric models. We will show that both
models yield similar wind speeds under a wide range of conditions.
A common characteristic of both exospheric models and kinetic
simulations is that, unlike fluid models, no assumption on energy
transport has to be made: the heat flux is completely
self-consistent. A detailed comparison of the results from both
exospheric models and collisional kinetic models will be presented
and discussed.

\section{Exospheric models and kinetic simulations} \label{models}

In exospheric or kinetic collisionless models of stellar
atmospheres, the plasma is assumed to be completely collisionless
beyond a given altitude, called the exobase. In principle, the
collisionless nature of the plasma above the exobase allows to
compute, for each particle species, the VDF at any arbitrary
height as a function of the VDF at the exobase, by means of
Liouville's theorem and the requirement of energy and magnetic
moment conservation. However, the task is not trivial because the
electric field profile needed to ensure local quasi-neutrality and
zero current is an unknown of the problem that has to be
determined self-consistently. The electric field arises because of
the small electron-to-proton mass ratio that makes it easier for
an electron, compared to a proton of same energy, to escape from
the star. In short, the electric force must be directed toward the
star for the electrons and away from the star for the protons.
This field is thus responsible for the strong outward acceleration
of the protons (see e.g. \citet{mak97a}).

The first models of this kind \citep{joc70, lem71} were based on
Maxwellian VDFs for the electrons and were unable to reproduce the
observed velocities of the fast solar wind unless extremely high,
and unrealistic, coronal temperature were assumed. Years later, it
became possible to reproduce the high speeds of the fast solar
wind by assuming Kappa distributions at the exobase
\citep{mak97a}; this is because the suprathermal electrons tend to
increase the flux of escaping electrons and therefore produce a
larger accelerating electric potential for the protons. These
early models assumed that the total proton potential energy
(gravitational + electrostatic) is a monotonic decreasing function
of the radial distance to the star. As a consequence, the exobase
was implicitly assumed to be located close to the
subsonic-supersonic transition level.

This model has been generalized by \citet{lam03} and \citet{zou04}
by relaxing the requirement of the proton potential energy being
monotonic. These authors found complete transonic solutions
describing both the subsonic and the supersonic regimes of the
fast solar wind. The basic outcome is a high value of the terminal
bulk speed (700-800 km s$^{-1}$), compatible with observed fast
solar wind speeds, by assuming a Kappa VDF for the high energy
electrons at the exobase without any assumption on energy
transport. It is noteworthy that this result is not an artifact of
the use of Kappa functions. \citet{zou04} were able to obtain
similar results assuming a sum of two Maxwellians, which is the
most commonly used model to represent the electron VDFs in the
solar wind.

\citet{lan03} have presented self-consistent kinetic simulations
of a stationary solar type wind using Maxwellian VDFs for the
protons and the electrons. The model is spatially one dimensional
and spherical symmetric, but particles' velocities are three
dimensional. In order to allow for binary collisions, the
following rule has been introduced: two particles crossing each
other at relative velocity $u$ at a distance $r$ from the star may
undergo an isotropic elastic collision with probability $\propto
u^{-4}r^{-2}$. The $u^{-4}$ dependence of the collision
probability mimics the velocity dependence of the scattering cross
section for Coulomb collisions, whereas the $r^{-2}$ dependence
accounts for the spherical geometry of the problem. The transport
properties of such a plasma have been shown to be similar to those
of a Fokker-Planck plasma \citep{pan01, lan01}. These kinetic
simulations have shown that the existence of a transonic wind
requires a minimum collisionality near the sonic point. In other
words, the coronal density must exceed a threshold density for the
wind acceleration to be sufficiently strong for the distant wind
to be supersonic. It was also shown that the electron heat flux
departs from the classical value \citep{spi53} in most of the
acceleration region. In the next section we present new results
from this model using Kappa VDFs for the electrons and the real
value of the proton to electron mass ratio, unlike \citet{lan03}
who used a reduced mass ratio of 400 for computational reasons.

\section{Results} \label{results}

In both the exospheric models and the kinetic simulations, we use
a Kappa VDF $f_{\kappa}(v)\propto (1+v^2/\kappa
v_{\kappa}^2)^{-(\kappa+1)}$. The equivalent Kappa temperature
$T_{\kappa}$ (defined from the second moment of the VDF, as the
ratio between pressure and density) is related to the thermal
speed $v_{\kappa}$ by $T_{\kappa}=(\kappa/(2\kappa-3))m_{\rm
e}v_{\kappa}^2/k_{\rm B}$, where $k_{\rm B}$ is the Boltzmann
constant and $m_{\rm e}$ the electron mass. For speeds smaller or
comparable to $v_{\kappa}$, the Kappa VDF is close to a Maxwellian
having the same most probable speed $v_{\kappa}$.  In contrast,
for $v\gg v_{\kappa}$, the Kappa VDF decreases with $v$ as a power
law $(f_{\kappa}\varpropto v^{-2(\kappa+1)})$. In the limit
$\kappa\rightarrow\infty$, $f_{\kappa}(v)$ reduces to a Maxwellian
VDF $(\propto e^{-v^2/v_{\kappa}^2})$. Note that when electron
distributions measured in the solar wind are fitted with Kappa
functions, the parameter $\kappa$ ranges from 2 to 5
\citep{mak97b}.

Besides the shape of the VDF, the physical state of the corona at
heliocentric distance $r_0$ (exobase) is characterized by a key
parameter, proportional to the ratio of the thermal energy of a
proton to its gravitational energy
\begin{equation}\label{parameteralphadefinition}
 \alpha\equiv
\frac{2v_{th0}^{2}}{v_{esc}^{2}}=\frac{2r_0k_{\rm B}T_0}{m_{\rm p}MG}\propto
r_0T_0
\end{equation}
where $M$ is the mass of the star and $T_0$ is the temperature at
the base of the wind, assumed for simplicity to be the same for
electrons and protons. In this case, the wind profiles can only
depend on $\alpha$ and on the shape of the VDF.

\begin{figure} \plotone{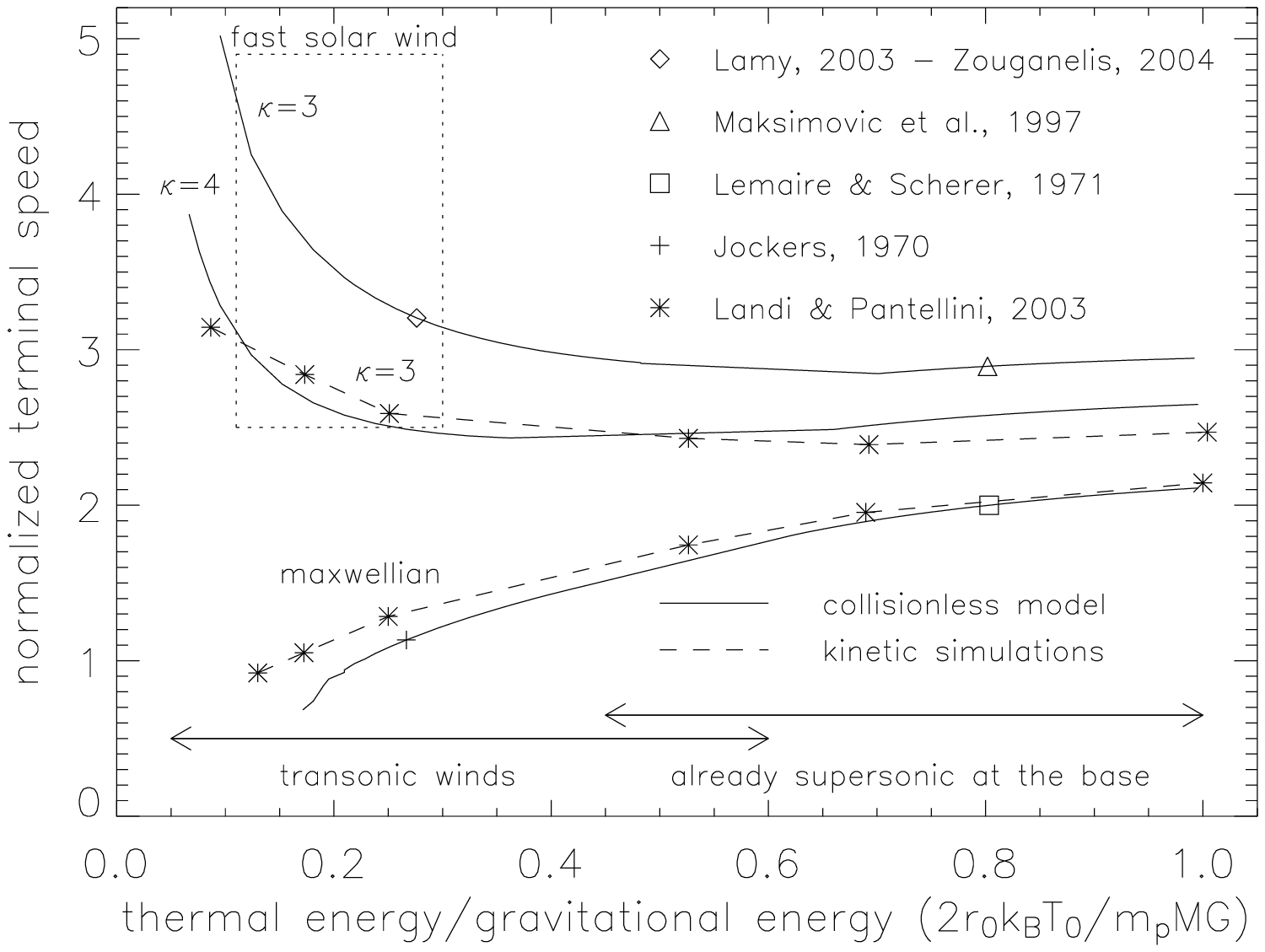} \caption{Terminal speed normalized to the
thermal speed either at the exobase (exospheric model) or the
lower boundary (kinetic model with collisions) as a function of
the dimensionless parameter $\alpha$ for different
models.\label{vitterm}}
\end{figure}

Figure $\ref{vitterm}$ summarizes our results. It shows the
terminal bulk speed normalized to the proton thermal speed at the
exobase as a function of $\alpha$ for different values of
$\kappa$. Results are shown for the exospheric model (full lines)
and kinetic simulations (dashed lines). The rectangle in the upper
left part of the figure covers the parameter space compatible with
observational data for the fast solar wind. When $\alpha$ is
large, the corona "explodes" and the wind starts at nearly
supersonic velocity. This is the case studied by \citet{lem71}
with a Maxwellian VDF and by \citet{mak97a} with a Kappa VDF. For
smaller values of $\alpha$, the gravitational force holds most of
the protons back, up to a radial distance where their potential
energy goes through a maximum and where the wind is already
supersonic. This case has been studied by \citet{joc70} with a
maxwellian VDF and by \citet{lam03} and \citet{zou04} with a Kappa
VDF, a sum of two maxwellians and a sum of a maxwellian and a
kappa function. The transition between a transonic and a
supersonic wind takes place at $\alpha=0.5$ in the traditional
Parker's model and at a slightly different value in the present
models.

Note that the points in Figure $\ref{vitterm}$ are \emph{based on}
the publications that they refer to, but are not the explicit
results of these publications. They are mainly illustrating the
different validity range of these works. All exospheric curves
(full lines) were obtained using the model by \citet{zou04}. The
simulation curves (dashed lines) were obtained using the model by
\citet{lan03} generalized to allow for Kappa electron VDFs as a
boundary condition at the lower boundary of the simulation domain.

Solutions of the exospheric problem are simple for the case where
the wind is already supersonic at the base, since in this case the
proton potential energy (gravitational + electrostatic) is a
monotonic decreasing function of the radial distance. Any proton
injected at the base is then doomed to escape to infinity, and
both the terminal speed and the asymptotic temperatures can be
calculated analytically \citep{mey98}. If the wind velocity at the
base is subsonic (the solar wind case), a local maximum appears in
the proton potential energy profile. Only a fraction of the
protons injected at the base are then able to escape to infinity
and the acceleration is found to be weaker than in the case of a
supersonic start at the base, at least for Maxwellian electron
VDFs \citep{zou04}.

When Kappa VDFs are used for the electrons, a larger acceleration
is attained as we can see in the curves based on \citet{mak97a}
and \citet{lam03}-\citet{zou04} for $\kappa=3$. For small values
of $\alpha$ (subsonic at the exobase), the speed increases as
$\alpha$ decreases, which means that the acceleration can be
stronger for a lower temperature. This basic difference from the
Maxwellian case is presumably due to the fact that the
acceleration is mainly sustained by the excess of suprathermal
electrons of the Kappa distribution and less dependent on the
thermal energy.

The dashed curves in Figure $\ref{vitterm}$ stem from kinetic
simulations with collisions. When Maxwellian electron VDFs are
injected at the lower boundary of the simulation, the results are
remarkably similar to those obtained using Maxwellian
distributions in the exospheric model. This means that neglecting
collisions in the exospheric models has no significant
consequences on the terminal bulk speed. When injecting Kappa
electron VDFs at the lower boundary, the curves from the kinetic
simulations with collisions are slightly different from those
obtained with the corresponding exospheric model, but the
qualitative behavior is similar (normalized terminal speed
decreasing with $\alpha$). In this case there is always some value
of $\kappa$ for the exospheric model giving the same results, in a
large range of $\alpha$, as the kinetic simulations with a lower
$\kappa$. As we can see, the kinetic simulations with $\kappa=3$
give almost the same results as the exospheric model for
$\kappa=4$ for fast solar wind compatible parameters (although
this does not imply a general rule). This suggests that collisions
tend to reduce the wind acceleration for a given kappa
distribution.

\begin{figure} \plotone{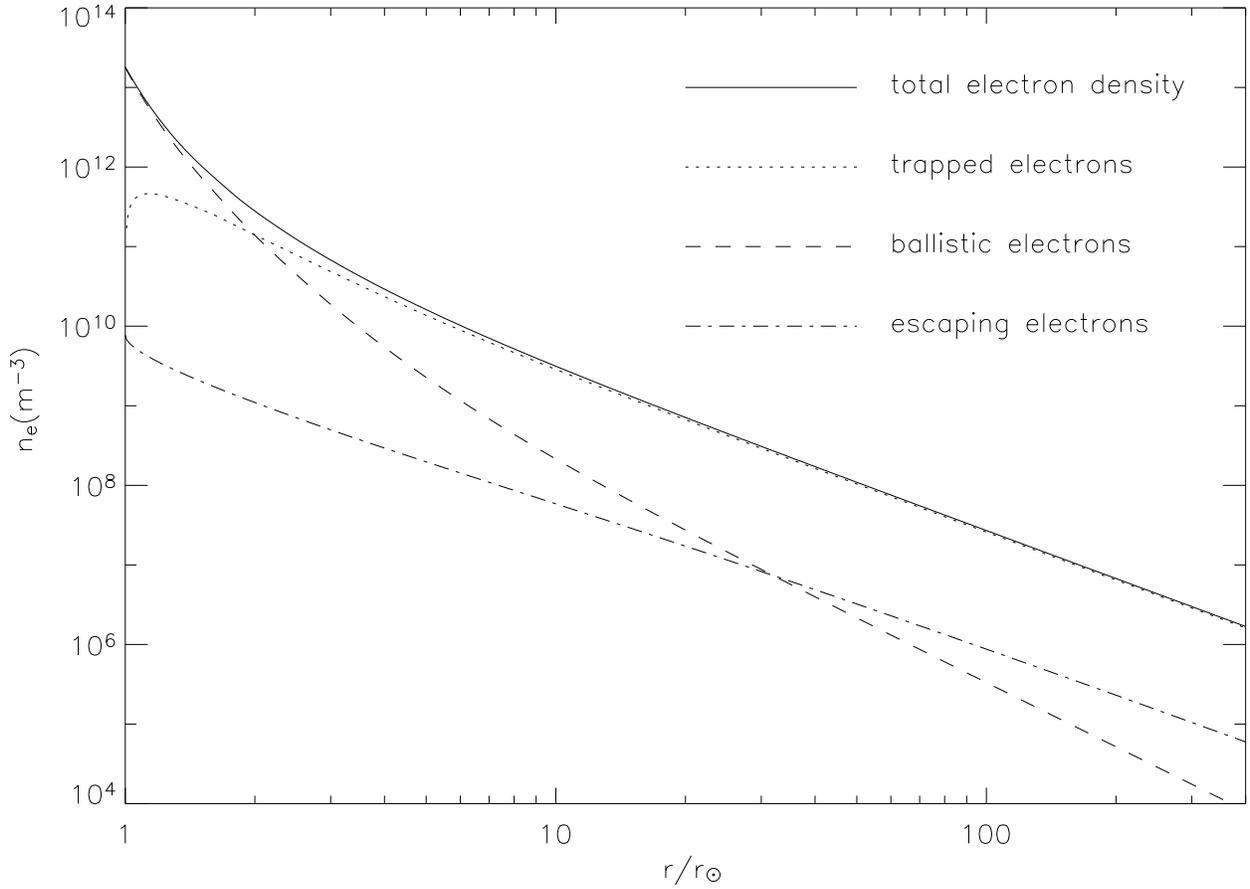} \caption{Electron density profile (solid line) for $\kappa=2.5$. The other lines show the contributions of the different electron
populations. Note that trapped electrons (dotted line) become
predominant beyond a few solar radii. \label{densepopul}}
\end{figure}

However, the agreement between exospheric collisionless models and
kinetic simulations with collisions is rather surprising. Indeed,
the main criticism usually raised against exospheric models is
their neglect of collisions. The relative agreement between
exospheric model and kinetic model including collisions may be
explained by the presence of trapped electrons in exospheric
models. These electrons do not have enough energy to escape from
the Sun and their inclination to the magnetic field lines is large
enough that they are reflected by the magnetic mirror force before
reaching the exobase. Trapped particles do not therefore exist at
the exobase, despite the fact that they rapidly become the
dominant component of the total electron density on the way from
the exobase to the maximum of the proton potential energy and
beyond. Figure $\ref{densepopul}$ shows that at large radial
distances, the trapped electrons represent more than $90\%$ of the
total electron density.

When completely withdrawing the trapped electrons from exospheric
models, no supersonic wind solution can be found numerically. This
is presumably due to the electron density being too small to
ensure local plasma neutrality together with a reasonable
configuration of the proton potential energy. In other words, if
all electrons do escape (except the small population of ballistic
ones, falling back to the exobase), quasi-neutrality and zero
electric current tend to become incompatible requirements of the
model.

This is in agreement with kinetic simulations showing that
collisions are necessary to accelerate the wind to supersonic
velocities. Collisions are responsible for the transformation of
ballistic particles into trapped ones. In exospheric models,
trapped particles were historically added in order to avoid
discontinuities in the VDFs at the interface between trapped and
untrapped orbits in phase space, but they seem to be crucial for
the acceleration of the wind to supersonic velocities. Their
presence in exospheric models implies that the latter are not
collisionless in a rigorous sense.

Note that a similar problem arises in the environment of space
probes \citep{laf73}. In that case, the small size of the probe
makes the medium both fully collisionless (there are no trapped
particles) and non neutral; there is a non zero space charge
albeit not exactly the canonical Debye sheath (see \citet{mey93}).
In contrast, in the solar wind case, the plasma has to be neutral
because the scales are much greater than the Debye length and for
this reason, the presence of trapped electrons is essential.

In simulations, collisions have been seen to serve to convert the
electron heat flux into plasma bulk energy. We have compared the
heat flux given by exospheric models and by kinetic simulations
and found a qualitative agreement. As was pointed out by
\citet{lan03}, the non classical term $q_{\rm NC}\propto
\frac{3}{2}nvk_{\rm B}T_{\rm e}$ of the electron heat flux
introduced by \citet{hol74} dominates the classical Spitzer-Harm
term for the supersonic wind. With both the exospheric model and
the simulations, we find a heat flux still several times greater
than the above value. This suggests that the classical formulation
is not the relevant one for such a semi-collisional medium, which
is not surprising as the Spitzer-Harm term was calculated upon the
assumption of a collision dominated plasma (Knudsen number much
smaller than unity).

\section{Conclusions} \label{conclusions}

Exospheric models are aimed at explaining the strong acceleration
of the fast solar wind in a self-consistent way with a minimum
number of assumptions. In particular, no assumption on how energy
is transported through the acceleration region needs to be
included in the model. The model is admittedly over simple in that
it is one-dimensional, time stationary, collisionless, and by
construction free of any wave activity. However, the basic
ingredients for powering stellar winds appear to be present,
suggesting that propulsion by plasma waves is not necessarily
needed to produce powerful transonic winds. Over the last three
decades various exospheric models were able to reproduce both the
slow and the fast solar wind, though in restricted wind regimes
only. For the first time, we generalize all previous models to a
much wider range of parameters by varying both the temperature and
the abundance of suprathermal particles. The generalization covers
a large class of coronal conditions, including the solar corona
case. For high temperature coronas or large stellar radii or small
stellar mass, the corona "explodes" and the wind starts supersonic
at the exobase (non solar case). The solar case is different as
the solar wind starts subsonic and becomes supersonic beyond a
distance of some solar radii. Treating this case makes exospheric
models much more complicated than before (an accurate numerical
description has been recently given by \citet{zou04}).

We have also compared these models with kinetic simulations that
include Coulomb-like collisions. These simulations have been made
for the first time using non-maxwellian functions. Rather
unexpectedly, the results of exospheric models and kinetic
simulations are in good agreement despite the wide difference in
both the physics (collisionless versus collisional) and the
methodology. The agreement between the two approaches is likely
due to the fact that a small amount of collisionality is implicit
in exospheric models in that particle trajectories which are not
accessible from the exobase are populated "by hand". The existence
of trapped electrons in exospheric models is a necessary condition
for the wind to be supersonic, just as collisions in kinetic
simulations are necessary to produce a supersonic wind.

Exospheric models are able to reproduce the strong acceleration of
the fast solar wind from the subsonic to the supersonic regime,
provided the electron VDF has a suprathermal tail. The similarity
of results from exospheric models and kinetic simulations with
collisions suggests that the main role of collisions is to feed
particles into trajectories that are not accessible from the
exobase (the trapped particle trajectories). Indeed, a small
amount of collisions is implicitly included in exospheric models
through the "by hand" populating of the trapped particles
populations. Even the high terminal speeds obtained in exospheric
models do not seem to be a consequence of the collisionless nature
of these models, the main reason for the strong acceleration being
the presence of suprathermal electrons (e.g. kappa distribution or
a sum of two maxwellians). Suprathermal electrons are found to
collide very little because of the $v^{4}$ dependence of the
collisional mean-free path. Collisions can therefore modify the
shape of the VDF at low energy, but the high energy suprathermal
tails are basically unaffected by collisions and so is the overall
wind acceleration. Despite their intrinsic limitations, exospheric
models are found to be a very convenient tool to explore the
physics of weakly collisional solar-type winds.

\clearpage


\clearpage
\clearpage

\end{document}